\documentclass[10pt]{iopart}
\expandafter\let\csname equation*\endcsname=\relax 
\expandafter\let\csname endequation*\endcsname=\relax 

\makeatletter
\def\ps@pprintTitle{%
 \let\@oddhead\@empty
 \let\@evenhead\@empty
 \def\@oddfoot{}%
 \let\@evenfoot\@oddfoot}
\makeatother
\renewcommand{\thetable}{\Roman{table}}
\usepackage{csquotes}
\usepackage{graphicx,color}
\usepackage{float}
\usepackage{amsmath}
\usepackage{amssymb}
\usepackage{multirow}
\usepackage{dsfont}
\usepackage{adjustbox}
\usepackage{pdflscape}
\usepackage{wasysym}
\usepackage{siunitx}
\usepackage{physics}
\usepackage[percent]{overpic}
\usepackage[dvipsnames]{xcolor}
\usepackage[table]{xcolor}
\usepackage[normalem]{ulem}
\usepackage{subcaption} 
\newcommand{\DREAM}{\textsc{Dream}}

\begin{document}

\title{Thermal modeling of runaway electron induced damage in the SPARC tokamak}
\author{T. Rizzi$^{a}$, K. Paschalidis$^{a}$, S. Ratynskaia$^{a}$, P. Tolias$^{a}$, I.~Ekmark$^{b}$, M.~Hoppe$^{a}$, R.A.~Tinguely$^{c}$, A.~Feyrer$^{c}$, T.~Looby$^{d}$}
\address{$^a$ Department of Electromagnetic Engineering and Plasma Physics, KTH Royal Institute of Technology, Stockholm SE-100 44, Sweden \\
$^b$~Department of Physics, Chalmers Universiy of Technology, Gothenburg SE-412 96, Sweden \\
$^c$~Plasma Science and Fusion Center, Massachusetts Institute of Technology, Cambridge, MA, USA 02139 \\
$^d$~Commonwealth Fusion Systems, Devens, MA, USA 01434}
\begin{abstract}
\noindent The integrity of plasma-facing components (PFCs) in tokamaks is critically challenged by transient events such as runaway electron (RE) impacts. We report the first systematic analysis of the thermal damage to tungsten-based PFC tiles comprising the SPARC outboard off-midplane limiters that is induced by RE beams formed during vertical displacement events. Parametric scans in RE impacting characteristics as well as energy-pitch distribution functions from the \DREAM{} code are employed for calculations of the volumetric heat loads. A realistic panel design is adopted to enhance the fidelity of the thermal analysis. The PFC thermal responses are compared in terms of in-depth temperature profiles and damage characteristics, such as melt depth and vaporization losses. 
\end{abstract}
\maketitle

\section{Introduction}\label{sec:introduction}

In ITER and future fusion reactors, runaway electrons (REs) constitute a major threat to the integrity of the plasma facing components (PFCs)\,\cite{Pitts_2025,Ratynskaia_2025b,Krieger_2025,Sweeney_2026}. REs can lead to deep volumetric melting and to thermal-stress driven explosions that should severely reduce the wall lifetime as well as to elevated temperatures at the bond interface with the cooling substrate that could even cause rupture and coolant leaks\,\cite{Pitts_2025,Ratynskaia_2025b,Krieger_2025}. The realistic modeling of the thermomechanical response of PFCs to RE incidence involves a chain of complex codes that address the multifaceted physics of the problem which range from RE formation, transport and impact on the wall to the nonlinear coupling between the high strain rates, the extreme temperatures and the high stresses developed inside the material\,\cite{Ratynskaia_2025b}. 

Recently, the first controlled experiments dedicated to RE-induced damage on graphite samples have been carried out in DIII-D\,\cite{Hollmann2025,Hollmann_2025b} leading to rapid developments in the thermomechanical modeling of the failure of brittle materials subject to RE incidence\,\cite{Ratynskaia_2025,Rizzi_2025a}. These were soon followed by the first controlled experiments dedicated to RE impacts on tungsten (W) samples that were carried out in April 2025 in the WEST tokamak\,\cite{Corre_2025}. These controlled experiments confirmed the empirical observation that the volumetric heating by REs triggers PFC explosions which are accompanied by the release of fast debris, consistently documented in accidental RE-PFC interaction events in many contemporary tokamaks\,\cite{Ratynskaia_2025b,De_Angeli_2023}. Rigorous analysis of such explosive events requires simulations of the full W thermomechanical response, including viscoplastic effects, internal phase transitions and compressible hydrodynamic flows, which constitutes an ongoing effort\,\cite{Ratynskaia_2025b}. 

In this investigation, we present a first exploration of RE-induced damage on realistic, W-based PFCs in the SPARC tokamak, a compact, high magnetic field device currently under assembly by Commonwealth Fusion Systems. Due to SPARC's high plasma current of $I_\mathrm{P} = 8.7~\mathrm{MA}$, a significant plasma-to-RE current conversion (${>}50\%$) is anticipated during unmitigated disruptions \cite{Sweeney_2020,Tinguely2021,Datta2025}. The REs can lose confinement and impact the first wall in many ways (drift orbits, stochastic fields, and more), but only one scenario is evaluated here: We consider the ``scraping off'' of REs during a cold vertical displacement event (VDE) as modeled with the 3D nonlinear MHD code M3DC1 \cite{Jardin_2012,Datta2026}. In this case, the plasma contacts the outboard off-midplane limiter. The PFC tiles at this location and the local magnetic field are assumed to be constant in time and serve as the basis for the following analysis.

Here, we employ a three-stage, one-way coupled workflow that was recently developed for the predictive modeling of the ITER W first wall thermal response to RE beams\,\cite{Ratynskaia_2025c}. For the ITER predictions, the first stage that concerns the RE loading prescription was based on DINA\,\cite{Pitts_2025} and JOREK\,\cite{Bergström_2024} simulations, while, for the SPARC predictions, we rely on parametric scans as well as simulations with the \DREAM{} code\,\cite{Hoppe_2021}. The second stage, which concerns Monte Carlo simulations of the RE volumetric energy deposition by the Geant4 code\,\cite{Geant4_2003,Allison_2006,Allison_2016}, and the third stage, which concerns heat transfer simulations by the MEMENTO code\,\cite{Paschalidis2023,Paschalidis2024}, remain the same as in Ref.\cite{Ratynskaia_2025c}.

\section{Problem postulation}


\subsection{Geometry}

\begin{figure}[!h]
    \centering
    \begin{overpic}[width=0.49\linewidth]{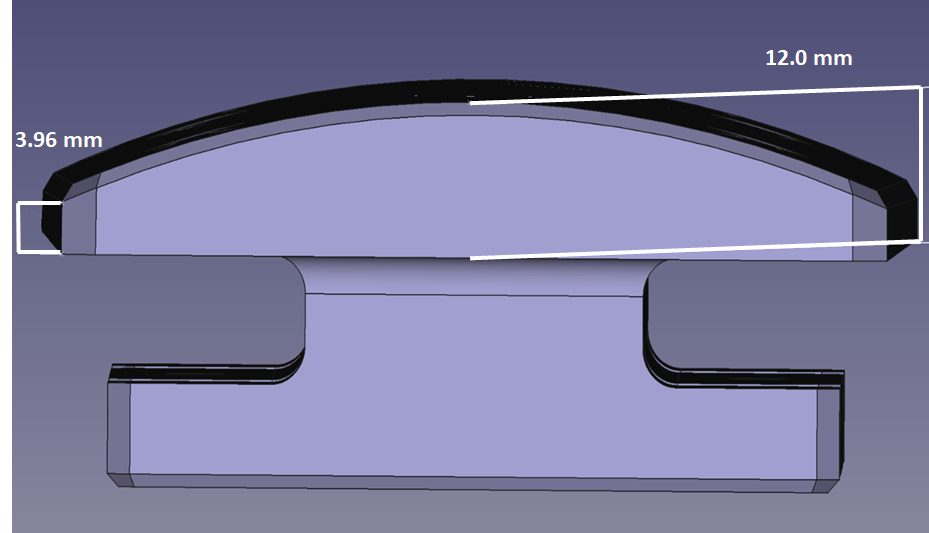}
        \put(2,48){\textcolor{white}{(a)}}
    \end{overpic}
    \hspace{0.1em}
    \begin{overpic}[width=0.44\linewidth]{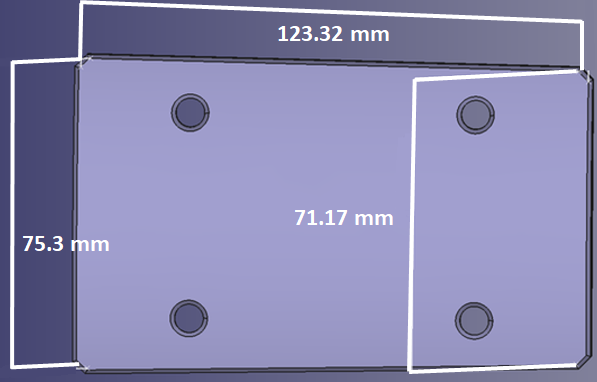}
        \put(2,57){\textcolor{white}{(b)}}
    \end{overpic}
    \caption{Geometry of the panel. View from the side (a) and the top (b).}
    \label{fig:panel_geometry}
\end{figure}

The focus here is on the outboard off-midplane limiters. The geometry of the plasma facing panels used in this work is shown in Fig.~\ref{fig:panel_geometry}. This realistic panel design was adopted to enhance the fidelity of the thermal analysis; however, it is important to note that this is a notional, not final, tile design.

\subsection{Material}

The panel is made up of W heavy alloy (WHA) with a composition of 97\%\,W, 2\%\,Ni, 1\%\,Fe. It is noted that the melting point of the Ni/Fe binder phase is much smaller than the melting point of W. In terms of the thermophysical behavior, a Ni/Fe depletion has been implicitly assumed due to the expected strong partial vaporization at elevated temperatures, which allowed us to use the well-known properties of pure W roughly above 2000\,K. Naturally, there could also be Ni/Fe segregation due to the enhanced mobility of the added elements\,\cite{Neu_2018}, but its effect on the W thermophysical properties cannot be quantified in a heuristic manner.

In the MEMENTO thermal response simulations, the temperature dependence of the WHA mass density was ignored and the room temperature value of 18.5\,g/cm$^3$ was used\,\cite{Neu_2018}. The temperature dependence of the specific isobaric heat capacity of the solid was constructed on the basis of WHA measurements from the room temperature up to 2100\,K, and of pure W measurements from 2300\,K up to the melting point of 3695\,K\,\cite{Tolias2017_1}. The transition between the two datasets was observed to be smooth without a need for artificial points. The synthetic dataset was accurately described by the analytical expression
\begin{align*}
c_{\mathrm{p}}(T)&=128.671+\frac{1.494901}{T}\times10^{3}+46.1738\times10^{-3}\times{T}\\&\quad-25.4798\times10^{-6}\times{T}^2+6.87537\times10^{-9}×T^3\,,
\end{align*}
with $c_{\mathrm{p}}$ measured in J/(kg\,K) and T in K. Moreover, an analytical expression for the temperature dependent thermal conductivity of the solid phase was constructed on the basis of WHA measurements from the room temperature up to 1750\,K\,\cite{Neu_2017}, which were least square fitted into an expression that can be extrapolated up to the W melting point of 3695\,K and reads as 
\begin{equation*}
k(T)=88.5798-\frac{3.396274}{T}\times10^{3}+5.70892\times10^{-3}\times{T}\,,
\end{equation*}
where $k$ is measured in W/(m\,K) and T in K. Extrapolation was favored over the generation of a synthetic WHA/W dataset (as for the heat capacity). It was preferred, since the WHA thermal conductivity is a monotonically increasing function of the temperature near 2000\,K\,\cite{Neu_2017}, whereas the pure W thermal conductivity is a monotonically decreasing function of the temperature near 2000\,K\,\cite{Tolias2017_1}, implying that a synthetic dataset would feature a spurious maximum. Nevertheless, the WHA fitting expression was confirmed to yield small high temperature deviations from pure W. Finally, the pure W recommendations of Ref.\cite{Tolias2017_1} were adopted for the specific isobaric heat capacity and thermal conductivity of the liquid phase that are applicable up to the normal boiling point and that are naively extrapolated beyond $6200\,$K. Thus, predictions of extreme temperature values are accompanied by additional uncontrolled uncertainties (which is particularly true for the thermal conductivity at temperatures beyond $10000\,$K) and must be viewed with caution\,\cite{Ratynskaia_2025c}. Pure W recommendations were employed for the temperature dependent vapor pressure\,\cite{Arblaster_2018} and hemispherical emissivity\,\cite{Forsythe_1934,Matsumoto_1999} as well as for the latent heats\,\cite{Tolias2017_1}.

In the Monte Carlo electron transport simulations, pure W has been assumed instead of WHA. In Geant4, alloys can be straightforwardly defined by specifying the elemental composition, which suffices to determine all the appropriate interaction cross-sections, and the mass density, which determines the number density of the scattering centers. However, the 3\% Ni-Fe content and roughly $1.05$ W/WHA mass density ratio imply minuscule differences in the energy deposition profiles. Thus, Geant4 simulations of RE transport into WHA were judged to be superfluous.

\subsection{Impact scenarios }

\subsubsection{Parametric scans.}

In the simplified input for the RE impact characteristics, the initial kinetic energy distribution is assumed to be mono-energetic with values of 0.5\,MeV, 1\,MeV, 10\,MeV and 50\,MeV, while the initial pitch angle - with respect to the local magnetic field - is assumed to be 0$^\circ$, 5$^\circ$ (${\sim}0.09~\mathrm{rad}$), and 10$^\circ$  (${\sim}0.17~\mathrm{rad}$). These individual particle energies and pitch angles were selected to cover the appropriate ranges for realistic RE distribution functions (shown in the next section).  The total energy loaded in a panel varies between 20\,kJ and 100\,kJ; for the case with 10\,MeV electrons, which corresponds to a RE current per panel of around 50\,kA and 250\,kA, respectively. Because the wetted areas of RE impacts are not well understood, these total energy and current values represent plausible scenarios for SPARC; for example, a 1\,MA RE beam could impact 4-20 tiles in this case. Additionally, these loadings show clear differences in melt characteristics (as will be described in the later sections). Finally, the RE loading time is assumed to be 1\,ms \& 10\,ms\; corresponding to either rapid loss over a duration shorter than the fastest expected current quench (${\sim}3.2\,\mathrm{ms}$) \cite{Sweeney_2020} or continuous loss over the course of a mitigated current quench (${\sim}5-10\,\mathrm{ms}$)\,\cite{Izzo_2025}. 

\begin{figure}
    \centering
    \begin{overpic}[width=0.49\linewidth]{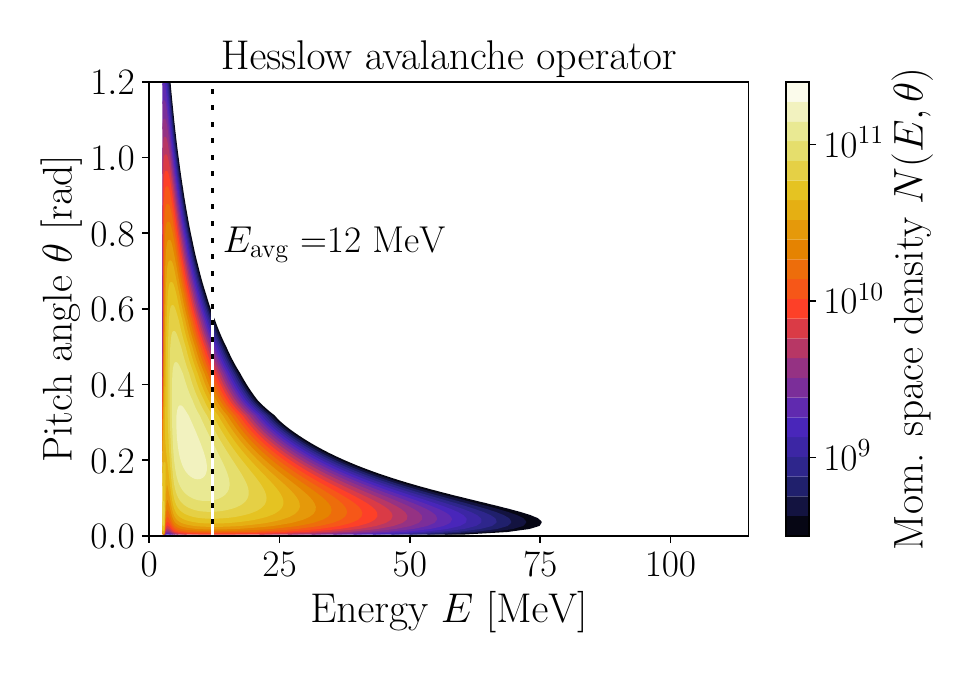}
        \put(14,65){(a)}
    \end{overpic}
    \begin{overpic}[width=0.49\linewidth]{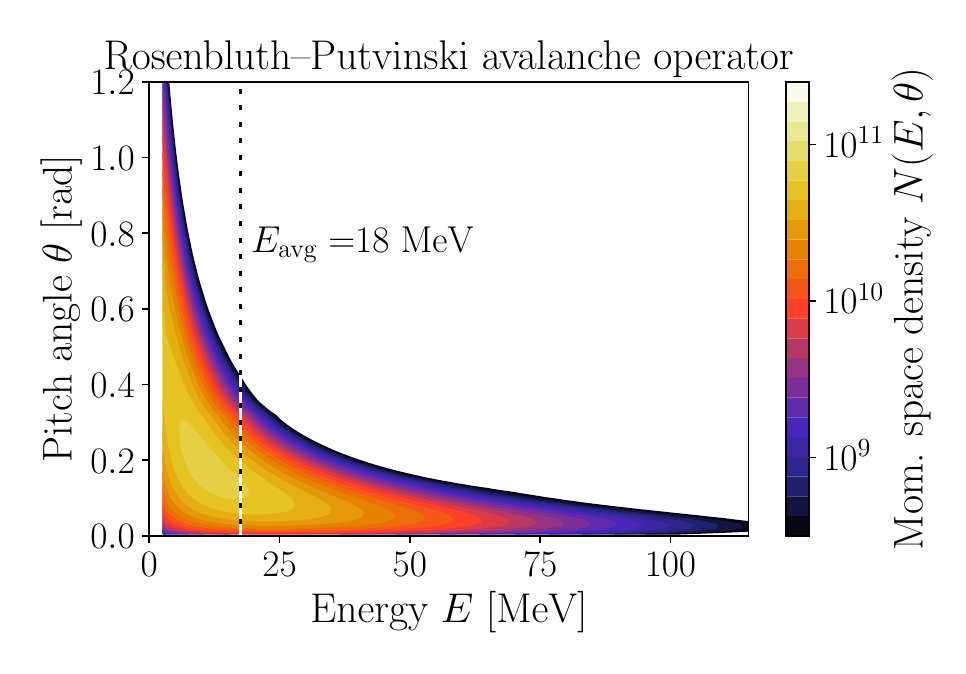}
        \put(3,65){(b)}
    \end{overpic}
    \caption{
        Runaway electron momentum space densities in energy-pitch angle space, $N(E,\theta)$: (a) calculated with the Hesslow fluid source term (\DREAM{} I), and (b) calculated with the Rosenbluth-Putvinski kinetic source term (\DREAM{} II).
    }
    \label{fig:distr_DREAM}
\end{figure}

\subsubsection{DREAM RE distributions.}

In a more realistic input for RE impact characteristics, \DREAM{} simulations have been employed that provide energy and pitch angle distribution functions. \DREAM{} is a fluid-kinetic simulation framework designed for RE modeling, particularly in tokamak disruption scenarios~\cite{Hoppe_2021}. Here we employ it to simulate a full SPARC disruption.

The distribution functions were obtained through \DREAM{} using the same baseline simulation setup as the one used and described in Ref.~\cite{Ekmark_2025}. 
Specifically, a deuterium-tritium plasma corresponding to the SPARC primary reference discharge was simulated, with a pre-disruption plasma current of \SI{8.7}{MA} and pre-disruption core temperature of \SI{22}{keV}. In Fig.\,1a of Ref.\cite{Ekmark_2025}, the radial profiles of the plasma current density, the plasma temperature and the electron density are plotted, while the equilibrium flux surfaces are plotted in Fig.\,1b.
The disruption is initialized by an instantaneous and uniformly distributed deposition of neon with a density of \SI{2.5e20}{m^{-3}}. We have simulated both the superthermal (electrons with momenta ${p\sim 0.001}$-$1\ m_{\rm e} c$) and the runaway electrons kinetically, while the post-disruption thermal electron population is treated as a fluid. Collisions involving the superthermal and/or runaway electrons are treated in the superthermal limit, meaning that the collisional friction force is taken to be $F_{\rm fr}\propto v^{-2}$. Since both the superthermal and runaway electron distributions are functions of pitch in our simulations, this choice of collision operator corresponds to the ``superthermal'' model as defined in Ref.~\cite{Hoppe_2021}.

To accurately represent the primary runaway generation, a finely spaced momentum grid is necessary at small momenta ($p\sim m_{\rm e} c$). At large momenta (${p\gg 1}\ m_{\rm e} c$), the dynamics are less sensitive to the momentum resolution, and a coarser grid spacing is advantageous for computational efficiency. For this reason, the momentum grid cell size is set to increase linearly with momentum. Additionally, for the superthermal electrons, the pitch angle grid spacing is chosen as to resolve the trapped passing boundary appropriately, while for the runaway electrons, the smallest pitch angles are more well resolved. 

Two different models for the avalanche generation are considered here: (I) avalanche multiplication with the Hesslow fluid source term~\cite{Hesslow_2019}, and the (II) kinetic Rosenbluth-Putvinski source term~\cite{Rosenbluth_1997}. In both these models, a kinetic equation of the form
\begin{equation}
    \frac{\partial f_{\rm re}}{\partial t} +
    \frac{\partial}{\partial\boldsymbol{p}}\cdot\left[\left(
       \boldsymbol{F}_E +
        \boldsymbol{F}_{\rm rad}\right) f_{\rm re}
    \right]
    = C\left[f_{\rm re}\right] + S_{\rm act} + S_{\rm ava},
\end{equation}
is solved for the runaway electrons, with $\boldsymbol{F}_E$ denoting the accelerating force due to the toroidal electric field, $\boldsymbol{F}_{\rm rad}$ the radiation-reaction force, $C[f_{\rm re}]$ the Fokker-Planck collision operator~\cite{Hoppe_2021,Decker2016}, $S_{\rm act}$ the generation of energetic electrons from tritium beta decay and Compton scattering~\cite{Ekmark_2024}. In scenario (I), the avalanche source term $S_{\rm ava}$ is taken to be
\begin{equation}
    S_{\rm ava} = \frac{V'}{\mathcal{V}'}\Gamma_{\rm ava}n_{\rm re}
    \delta\left(p-p_{\rm min}\right)\delta\left(1-\xi\right),
\end{equation}
where $V'$ denotes the configuration space Jacobian, $\mathcal{V}'$ the phase-space Jacobian, $\Gamma_{\rm ava}$ is the avalanche growth rate as given in Ref.~\cite{Hesslow_2019}, $n_{\rm re}=n_{\rm re}(r)$ is the runaway electron density, $\delta(x)$ is the Dirac delta function, $\xi=\cos{\theta}$ is the electron pitch, and $p_{\rm min}$ denotes the minimum momentum value on the runaway electron computational grid in \DREAM. In scenario (II), $S_{\rm ava}$ is instead taken to be of the Rosenbluth-Putvinski form~\cite{Rosenbluth_1997} with
\begin{equation}
    S_{\rm ava} = n_{\rm re}n_{\rm tot} cr_0^2
        \frac{\left\{B\delta\left(\xi-\xi^\star\right)\right\}}
        {\left\langle B\right\rangle}
        \frac{1}{p^2}
        \frac{\partial}{\partial p}\frac{1}{1-\gamma}.
\end{equation}
Here, $n_{\rm tot}$ is the combined density of free and bound electrons in the plasma, $c$ is the speed of light, $r_0$ the classical electron radius, $\gamma$ is the relativistic factor, and $\xi^\star=\sqrt{(\gamma-1)/(\gamma+1)}$ is the birth pitch, obtained from momentum conservation considerations. Curly braces $\{\cdot\}$ denote a bounce average while angle brackets $\langle\cdot\rangle$ denote a flux-surface average.

The difference in outcome between (I) and (II) lies in the energy distribution of the electrons which have undergone large-angle collisions. In (II), the incoming electrons are assumed to have infinite energy, and so the energy spectrum will extend to infinity with an exponentially decaying slope. This overestimates the extent of the energy spectrum and the maximum runaway energy. Since the majority of the secondary electrons will be generated near the critical momentum $p_{\rm c}$ for runaway acceleration, which is typically $p_{\rm c}\ll m_e c$, it might instead be more appropriate to use approach (I) and add all generated secondary runaways at an arbitrary small value of $p>p_{\rm c}$. This ensures that the maximum runaway energy is not overestimated, while maintaining the correct number of secondary runaways. The trade-off is that the pitch angle distribution of the newly born secondary runaways is inconsistent with the constraints introduced by momentum conservation in the knock-on collisions. However, considering that the electric field will quickly raise the parallel energy of these particles significantly, the exact birth pitch angle should not matter much for the final distribution function.

For the purposes of this paper, it is advantageous to define and use the runaway electron momentum space density
\begin{equation}
    N(E,\theta)=\int_0^a\dd r \mathcal{V}'\frac{\partial(p,E)}{\partial(\xi,\theta)}f_{\rm re}=
    \int_0^a\dd r \mathcal{V}'\frac{\sin\theta}{v}f_{\rm re},
\end{equation}
which is a function of kinetic energy $E$ and pitch angle $\theta$, defined such that $N_{\rm re}=\int_0^\pi\dd\theta\int_0^\infty \dd E\ N(E,\ \theta)$. In Fig.\,\ref{fig:distr_DREAM}, the momentum space density is plotted for the simulations corresponding to both avalanche operators. The reason why the maximum of the momentum space density is not located at $\theta=0$ is because the Jacobian $\partial(p,E)/\partial(\xi,\theta)$, which is already included in the definition of the momentum space density, contains a factor $\sin\theta$ due to $\mathrm{d}\xi/\mathrm{d}\theta$. 

From the RE momentum space density, we can calculate the total energy available in the RE beam. With $E_{\rm tot} = \iint N(\theta,E)E\,\mathrm{d}\theta\mathrm{d}E$, we obtain \SI{1.37}{MJ} and \SI{1.29}{MJ} for scenarios (I) and (II), respectively. The thermal response to the \DREAM{} input is evaluated assuming a loaded energy of \SI{100}{kJ}, which corresponds to an RE beam uniformly spreading over 14 and 13 identical tiles of the SPARC first wall, for scenarios (I) and (II), respectively.

From the distribution function we can also evaluate other important metrics for assessing the RE population. In both simulations, a runaway current plateau is established, of \SI{5.8}{MA} with model (I) and \SI{3.7}{MA} with (II). 
The total kinetic energy carried by the REs is \SI{1.37}{MJ} for model (I) and \SI{1.30}{MJ} for (II), while the average kinetic energy for REs in the distributions is \SI{12.1}{MeV} and \SI{17.5}{MeV}, respectively. 
Importantly, the runaway current and mean kinetic energy we obtain with model (II) is sensitive to certain numerical choices in the model. 
We have chosen a numerical configuration that promotes a higher mean energy, which corresponds to a lower plasma current, as this contrasts model (I) well, which instead has a slightly lower mean kinetic energy on the lower side, as discussed previously. 
That the runaway electrons in model (II) are more energetic, combined with the fact that they are fewer in number, explains why the two simulations predict a similar total kinetic energy of the runaway beam but different values of the runaway current. 


\section{Monte Carlo simulations of the energy deposition}\label{sec:experiment}

As detailed in Ref.\,\cite{Ratynskaia_2025c}, our Geant4 simulations for W have been benchmarked against MeV range calorimetric measurements\,\cite{Lockwood_1980}, keV range normal \& oblique electron backscattering experiments\,\cite{Bronshtein_1969,Gomati_2008,Hunger_1979,Reimer_2000} and MeV range normal electron backscattering measurements\,\cite{Tabata_1967} on high-Z refractory metals. Simulations followed the single scattering scheme implemented in the \textit{G4EmStandardPhysicsSS} library, whose modeling of the elastic scattering of electrons and positrons by neutral atoms is based on relativistic partial wave analysis for an effective interaction potential carried out with the ELSEPA code\,\cite{Salvat_2005,Salvat_2021}. Other electromagnetic processes were simulated with interaction models adopted from the PENELOPE\,\cite{Penelope} and Livermore\,\cite{Depaola_2006} libraries, while neutron production and transport were modeled using the G4NeutronHP package and the ENDF/B-VII.0 cross-section library\,\cite{Mancusi_2014}.

\subsection{Geant4 implementation of parametric scenarios}\label{sec:Geant4Impl}

\begin{figure}[!h]
    \centering
    \includegraphics[width=0.8\linewidth]{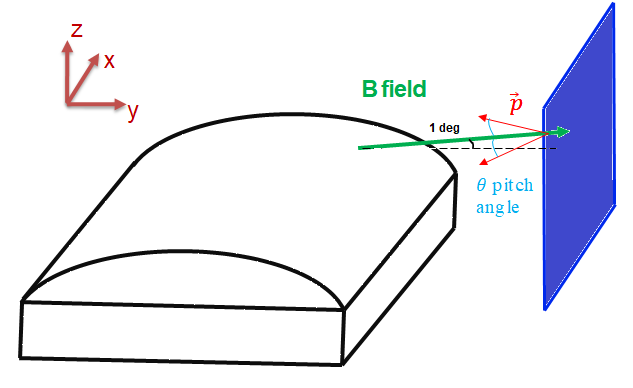}
    \caption{Geant4 set-up for the full panel simulations accounting for the panel curvature, varying magnetic field inclination angle and finite pitch angle. The direction of the incident REs and the magnetic field are indicated in red and green, respectively. Strength of magnetic field is $|\textbf{B}| = 13.2$\,T.}
    \label{fig:panel_geometry_Geant4}
\end{figure}

Since the acquisition of adequate Monte Carlo statistics uniformly across the entire 3D domain of the actual panel geometry (Fig.\ref{fig:panel_geometry}) is not computationally feasible, a two-step approach had to be adopted\,\cite{Ratynskaia_2025c}:

(i) \textit{Full panel simulations} with moderate particle statistics are first carried out to get a global picture as well as to capture the effects of the panel geometry, magnetic field topology and cascade particle transport. As illustrated in Fig.\,\ref{fig:panel_geometry_Geant4}, these Geant4 simulations are performed with a uniformly distributed particle source. The wetted area is obtained from the intersection of the B-field lines with the panel. The RE impact angles as well as the shadowing effect due to the curvature of the sample and the shallow magnetic field lines are reproduced in such simulations. Here, 10$^7$ electrons have been launched and a $\sim$\SI{100}{\micro\meter} mesh size has been employed for the evaluation of the energy deposition.

\begin{figure}[!h]
    \centering
    \begin{overpic}[width=0.8\linewidth]{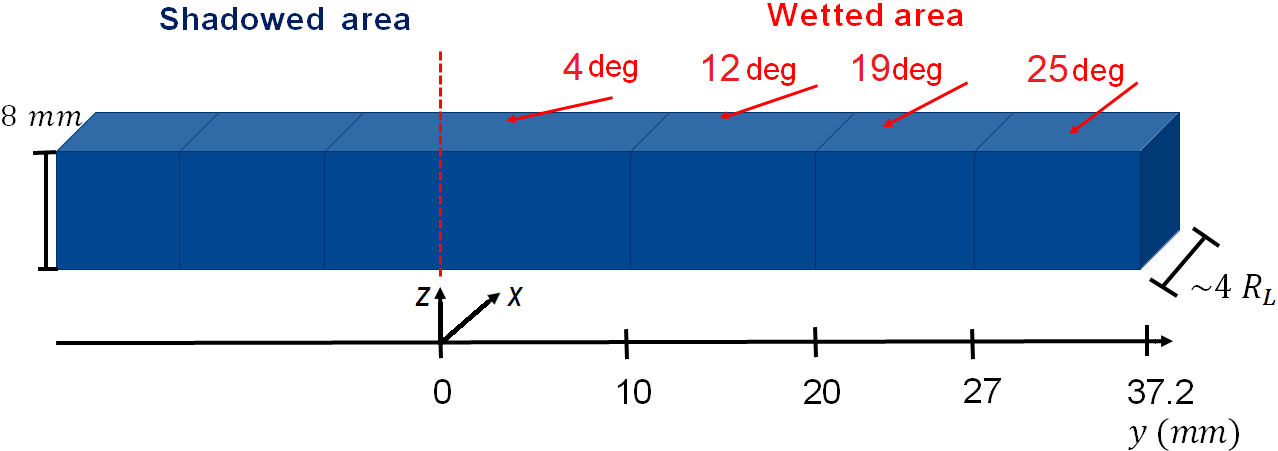}
    \put(36,19){\hbox{\kern1pt\textcolor{yellow}{\textbf{\tiny sec. 1} }}}
    \put(49,19){\hbox{\kern1pt\textcolor{yellow}{\textbf{\tiny sec. 2} }}}
    \put(63,19){\hbox{\kern1pt\textcolor{yellow}{\textbf{\tiny sec. 3} }}}
    \put(76,19){\hbox{\kern1pt\textcolor{yellow}{\textbf{\tiny sec. 4} }}}
    \end{overpic}
    \caption{Geant4 set-up for the simplified geometry simulations with the division into sectors 1-4. The red arrows indicate the direction of the incident RE beam.  For each sector, the average value of the magnetic field inclination angle $\alpha$ over the corresponding actual curved surface of the panel is provided.}
    \label{fig:Geant4_geometry_tile}
\end{figure}
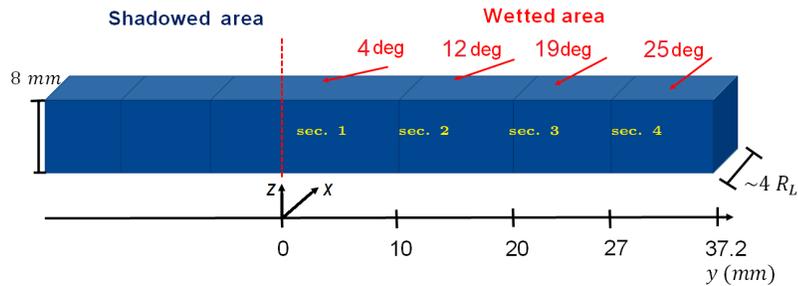

(ii)  \textit{Simplified geometry simulations} are carried out, characterized by high particle statistics and fine resolution, to yield highly accurate energy deposition maps. In particular, the curved panel is stretched to a slab that is truncated with respect to the x-axis to $4$ Larmor radii (due to the uniform source) sufficient to capture all returning backscattered electrons. The slab is divided into four sectors, labeled 1 to 4 as shown in Fig.\ref{fig:Geant4_geometry_tile}. The energy deposition map of a single-point particle source with 10$^6$ electrons is utilized to uniformly load each sector. Each single-point source launches electrons of a given pitch angle with respect to the magnetic field, whose inclination $\alpha$ (see Fig.\ref{fig:Geant4_geometry_tile}) is uniform within a sector and equal to an average value of the B-field incident angle over the corresponding actual curved surface of the panel. In addition, aiming to assess the statistical accuracy, tests were performed with increased statistics of 10$^7$ electrons resulting in a maximum discrepancy between the energy deposition profiles of merely 0.5\%. Finally, the four depth energy deposition profiles are normalized according to the value obtained from the full panel simulations. 

Given the reduced size of the simulation domain, higher resolution can be reached by implementing a non-uniform mesh along the in-depth dimension. In particular, the first near-surface layer of $1\,$mm has a resolution of \SI{20}{\micro\meter} that is relaxed to \SI{40}{\micro\meter} in the intermediate $4\,$mm layer and that is further relaxed to \SI{100}{\micro\meter} in the remaining $3\,$mm layer. We note that the \SI{20}{\micro\meter} resolution at the upper layer, where the steepest gradients are confined, was chosen with the thermal response in mind.  Essentially, a fraction of the characteristic length of heat diffusion over the loading time should define an adequate discretization\,\cite{Ratynskaia_2025c}.

It is noted that, in the cases with non-zero pitch angle, the randomization of the gyro-phase is also considered. For completeness, it is pointed out that the $50\,$MeV scenarios have been simulated with a slightly different domain compared to that shown in Fig.\ref{fig:panel_geometry_Geant4}. The reason is that at such high energies and for the steepest impact angle (sector 4), the RE beam deposits a significant amount of its energy much deeper than the 4\,mm depth which corresponds to the actual sample thickness towards the edges. Hence, the simulations for these scenarios were performed with a 4\,mm thick sector 4 to allow for correct energy normalization. 

\subsection{Geant4 implementation of DREAM RE distribution functions}

The main challenge in the Geant4 implementation of the RE distribution functions (Fig.~\ref{fig:distr_DREAM}) concerns the representation of every point of the two-dimensional energy-pitch phase space with sufficient statistics. The energy-pitch discretization from the \DREAM{} code is first re-interpolated into a uniform discretization of around $10^3$ $(E,\theta)$ pairs. This is again followed up by a two-step approach, similar to that described in Sec.\,\ref{sec:Geant4Impl}:

(i) \textit{Full panel simulations} are carried out with a uniformly distributed particle source identical for each $(E,\theta)$ pair, with $10^5$ statistics for a given pitch angle, including the gyro-phase randomization between 0 and 2$\pi$. The resulting energy deposition maps are then combined accounting for the probability density function at the specific point $(E,\theta)$.

(ii) \textit{Simplified geometry simulations} are carried out on a similar geometry to Fig.\ref{fig:Geant4_geometry_tile}, with an extension along the x-axis that is given by the Larmor radius of the highest energy point in the distribution function. For every $(E,\theta)$ pair, a single-point particle source with $10^5$ electrons is utilized to uniformly load each sector, again accounting for the gyro-phase randomization. Aiming to assess the statistical accuracy, simulations with $10^6$ particles per $(E,\theta)$ pair were also carried out, revealing a less than 3\% maximum deviation in the in-depth energy profiles. A weighted summation of the produced energy deposition profiles is calculated using the probability density function, which is re-normalized according to the energy fraction per section obtained from the full panel simulations.

\section{MEMENTO simulations of the thermal response}

The Geant4 energy density maps  (from the simplified geometry simulations with high particle statistics) define the volumetric heat source employed in the MEMENTO simulations. Translation of the Geant4 energy density to volumetric heat is done within the assumption that the RE impact conditions are uniform over the loading time, which is either 1\,ms or 10\,ms. 

The Geant4 simulations are based on the pristine geometry and provide the spatially dependent energy density maps. However, even when fully neglecting the mechanical response of the solid and the hydrodynamic response of the liquid, there is appreciable material erosion due to vaporization in case of high surface temperatures. Consequently, the naive loading of the Geant4 map without any translation to compensate for the vaporization would imply the loss of the energy deposited at the vaporized cells. In order to compensate, the heat source maps are shifted down (along the z-direction) to account for the fact that the free-surface position is evolving in time.

The energy density maps for the simulated scenarios do not feature gradients along the x-direction, which can be excluded from the thermal simulations. Moreover, the sector dimensions in the y-direction are much longer than the heat diffusion length in WHA even during $10\,$ms (\SI{500}{\micro\meter} versus $\,10\, \text{mm}$). Therefore, it would suffice to carry out four 1D heat transfer simulations, one for each sector. Herein, for the sake of simplicity in the MEMENTO implementation, we instead performed a 3D simulation on a full (stretched) panel domain as shown in Fig.\,\ref{fig:Geant4_geometry_tile}. Unlike Fig.\,\ref{fig:Geant4_geometry_tile}, there is no restriction that the width along the x direction should respect the Larmor radius. The discretization is very coarse along the x direction in order to reduce the computational time, while the cells are \SI{100}{\micro\meter} along the y direction and \SI{20}{\micro\meter} along the z direction aiming to resolve the steep temperature gradients. 

The boundary condition on the top surface is an inhomogeneous Neumann condition that features vaporization and thermal radiation cooling. Insulating boundary conditions are enforced on all other surfaces. Note that the insulating boundary inherently treats the fact that the actual sample is larger than the simulated sample in the y direction. Assuming that the loading is uniform, there are no temperature gradients along this direction, which is effectively equivalent to thermal insulation.  Moreover, the free surface is moving due to vaporization and the eroded cells are removed in the course of the simulation.

\section{Results}\label{sec:results}

\subsection{Simulations for simplified scenario}

\subsubsection{Energy distribution on the panel.} \label{sec:E_distr}

\begin{figure}[!h]
    \centering
    \includegraphics[width=0.8\linewidth]{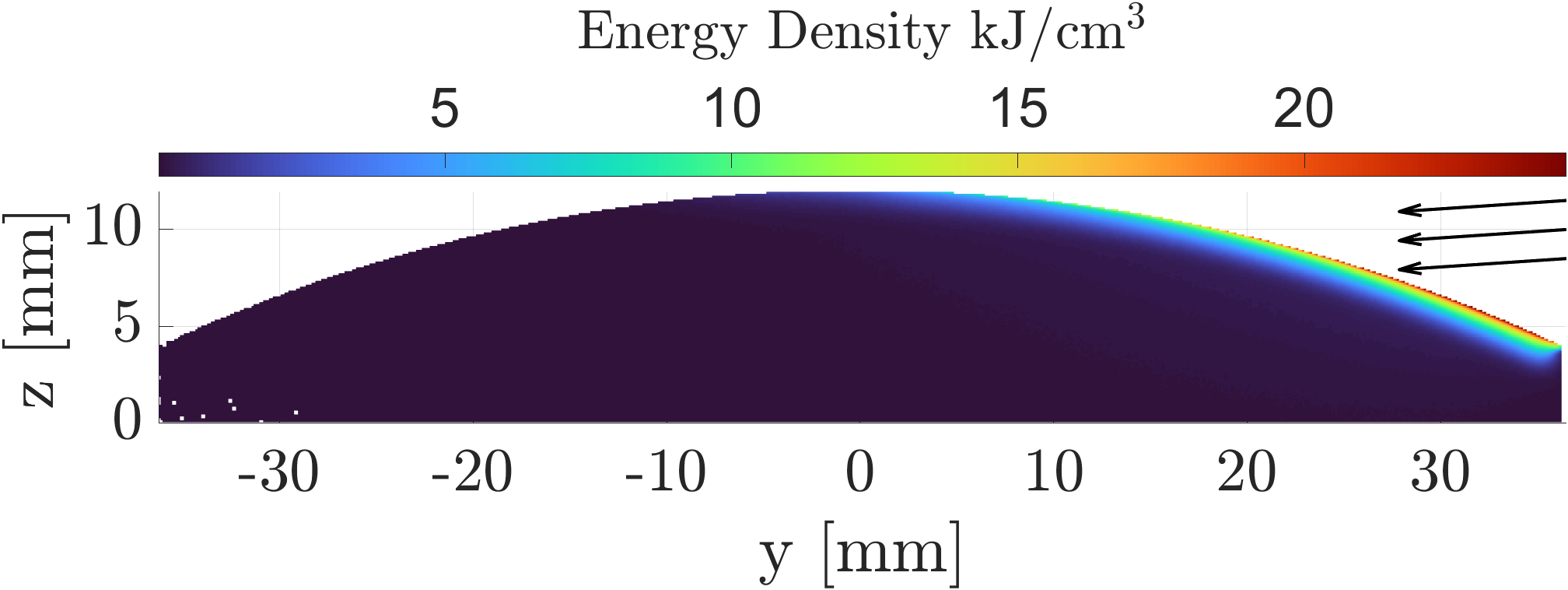}
    \caption{Results of full panel simulation for energy deposition by an RE beam of 10 MeV\,\&\,5${^\circ}$ pitch. Cross section view for normalization of 100 kJ. The incoming REs are indicated by black arrows. Note that shadowing from neighboring tiles is not considered.}
    \label{fig:full_panel}
\end{figure}

As aforementioned, the full panel Geant4 simulations allow the assessment of the energy received by each of the four sectors. Given the uniform electron source, the curved panel geometry implies that the lower side - sector 4 - should intersect most of the trajectories and, thus, receive the largest portion of the total energy loaded per panel (see Fig.\ref{fig:full_panel} as an example). Such a pure geometrical reasoning would result in an energy partitioning between the sectors which follows a sin$\alpha$ scaling. However, the full details of the energy deposition and the effect of the returning backscattered electrons are such that the energy deviates from pure geometrical estimates due to smearing out effects which become stronger with increasingly larger RE energy and Larmor radius. As a result, in all the scenarios with energies $\leq$10\,MeV, the deviations from the sin$\alpha$ scaling are within 15\% with sector 4 receiving about a factor of 3 times more energy compared to sector 1. On the other hand, in the scenario with energies of 50\,MeV, sector 4 receives only a factor of 1.2 times more energy compared to sector 1.

\subsubsection{Energy density distribution on the panel.}\label{sec:E_dens_distr}

\begin{figure}[!h]
    \centering
    \addtolength{\tabcolsep}{-9pt} 
    \begin{tabular}{cc}
        \subfloat{%
          \begin{overpic}[width = 2.5in]{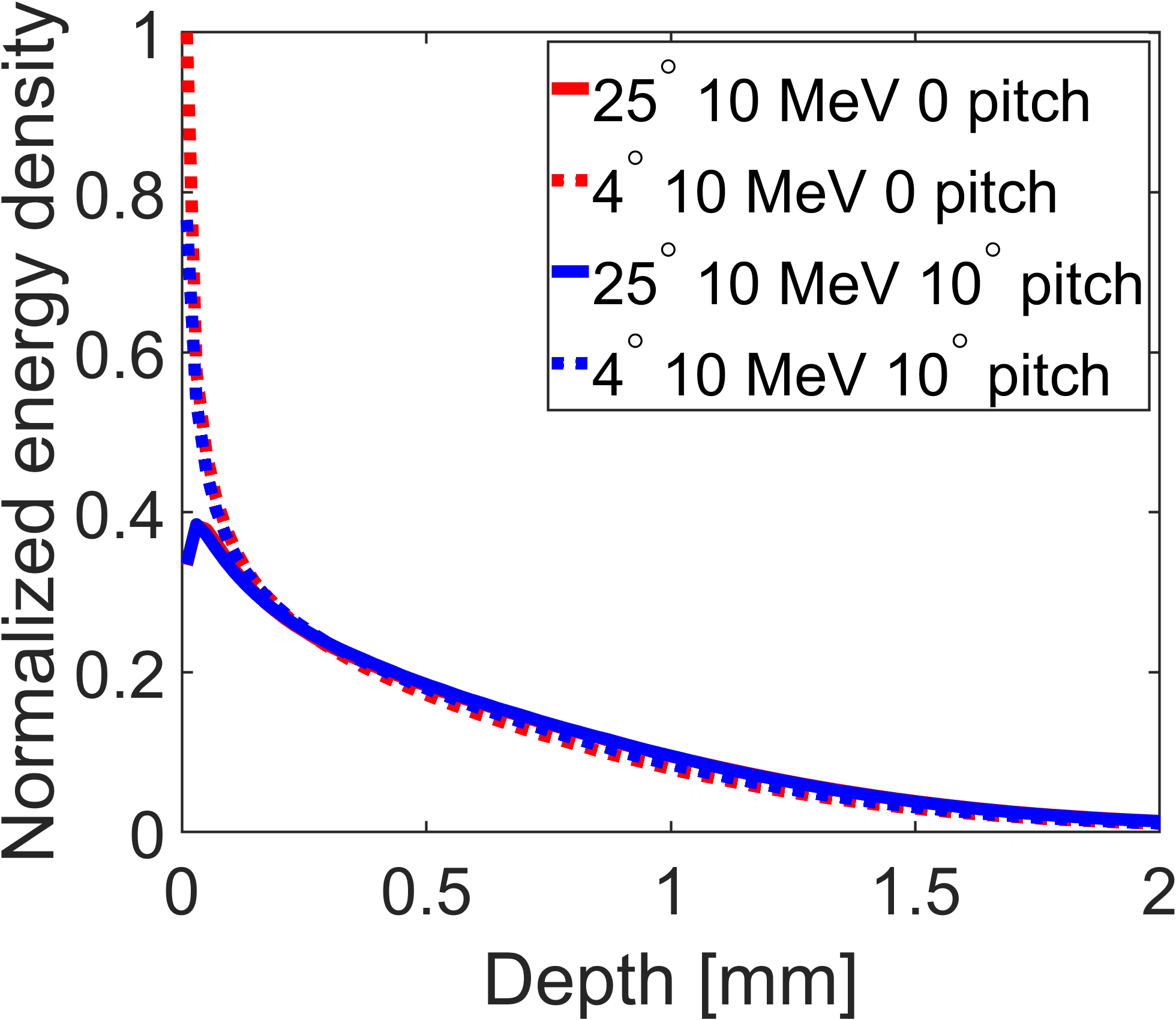}
          \put(80,30){\hbox{\kern3pt\textcolor{black}{\textbf{a)} }}}
          \end{overpic}
        }&
        \subfloat{%
          \begin{overpic}[width = 2.5in]{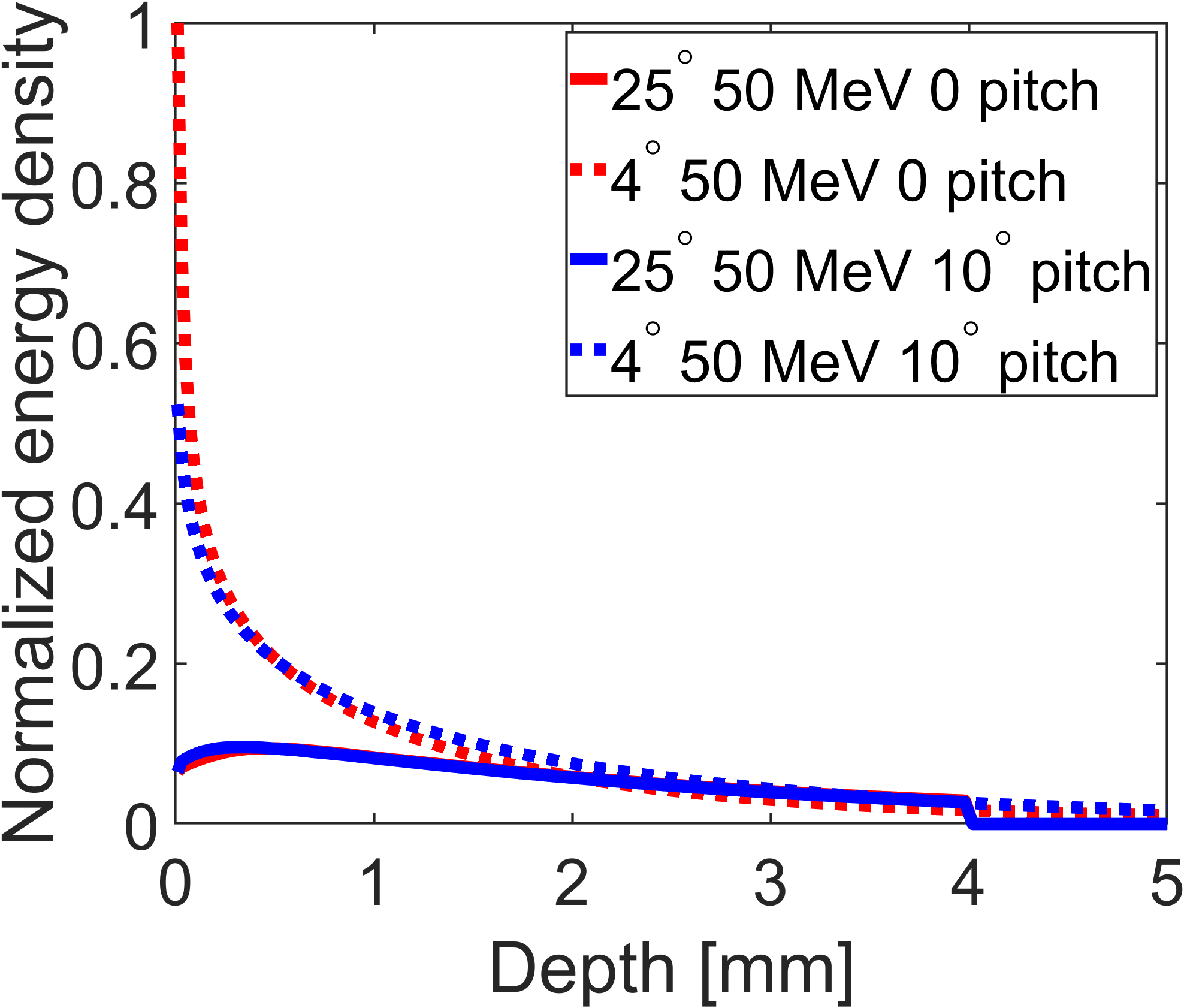}
          \put(80,30){\hbox{\kern3pt\textcolor{black}{\textbf{b)} }}}
          \end{overpic}
        }\\
        \subfloat{%
          \begin{overpic}[width = 2.5in]{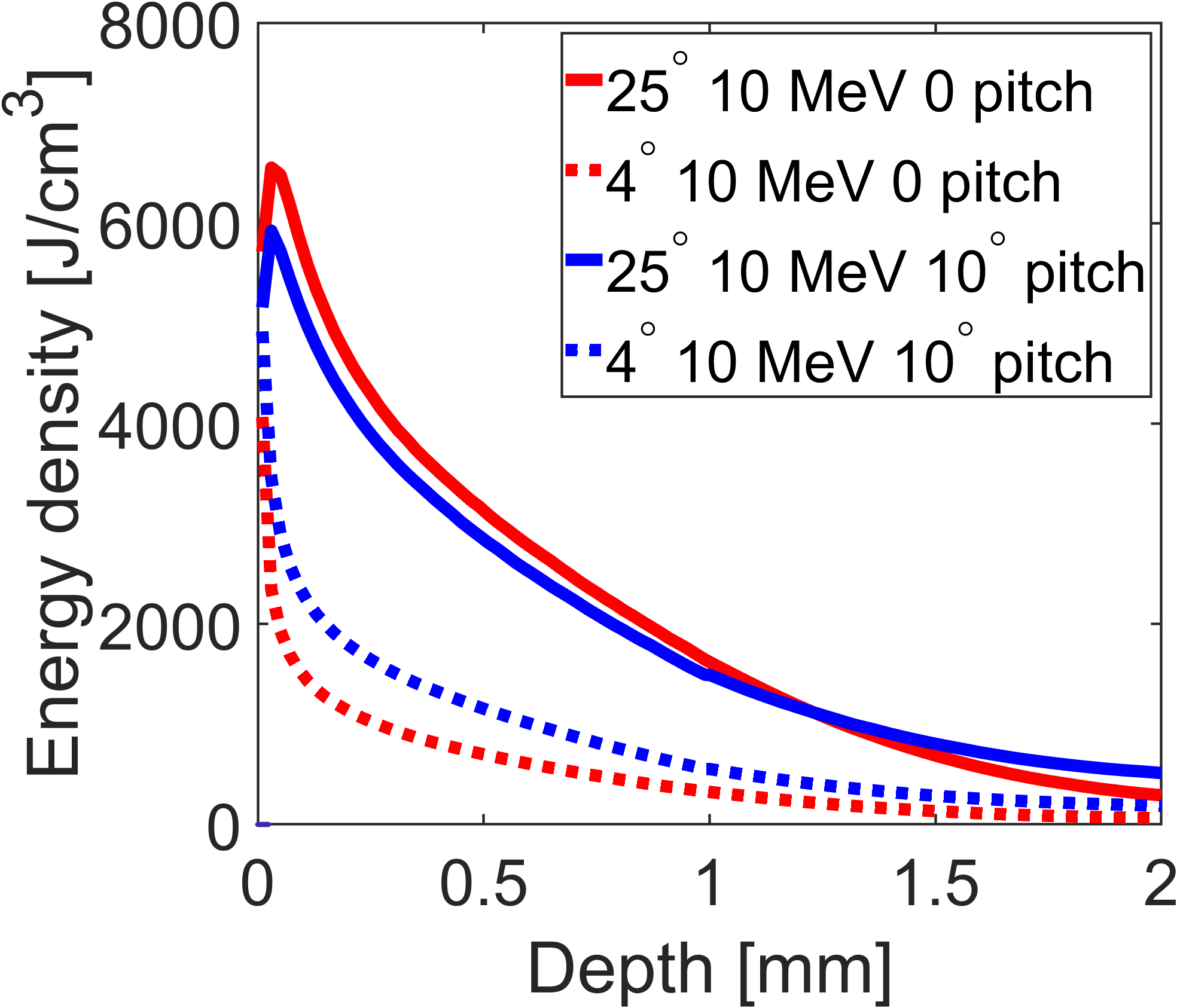}
          \put(80,30){\hbox{\kern3pt\textcolor{black}{\textbf{c)} }}}
          \end{overpic}
          }&
          \subfloat{%
          \begin{overpic}[width = 2.5in]{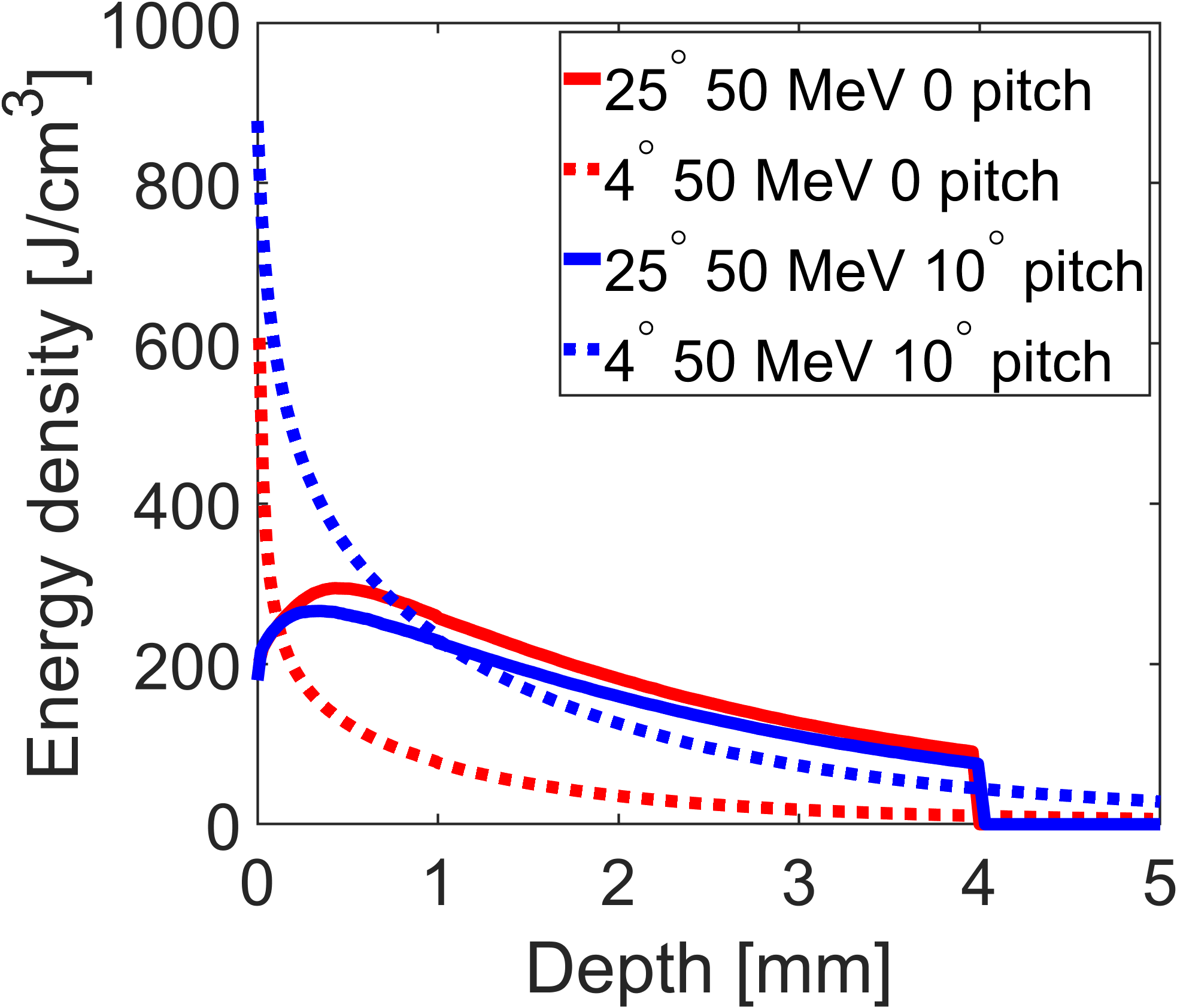}
          \put(80,30){\hbox{\kern3pt\textcolor{black}{\textbf{d)} }}}
          \end{overpic}
          }
    \end{tabular}
    \addtolength{\tabcolsep}{9pt} 
    \caption{In depth energy density profiles for sectors 1 (4$^\circ$) and 4 (25$^\circ$), with initial kinetic energies of 10 MeV and 50\,MeV, initial pitch angles of 0$^\circ$ and 10$^\circ$. In (a-b), the curves are normalized to the same energy deposited in the corresponding section, in arbitrary units. In (c-d),  normalization to 1\,kJ in the panel according to full panel simulations. We note that the profiles for 50\,MeV on sector 4 are obtained on a geometry with reduced thickness of 4\,mm as explained in Sec.\,\ref{sec:Geant4Impl}.}
    \label{fig:energy_densities}
\end{figure}

If the differences in the energy loaded per sector had been disregarded, then the highest energy density values would have always been encountered in sector 1, which is characterized by the most shallow impact angle. However, this sector receives the smallest energy (see above), hence higher energy density values can be encountered in other sectors. General conclusions can be drawn: (i) for energies up to and including 10\,MeV, the highest energy density values concern sector 4, (ii) for 50\,MeV the highest energy density is loaded on sector 2 or 1, depending on the pitch angle.


Fig.\ref{fig:energy_densities} presents in-depth energy density profiles for the RE initial electron kinetic energies of 10\,MeV (left column) and 50\,MeV (right column). The plots compare the pitch angle cases of 0$^\circ$ (red) and 10$^\circ$ (blue) for sectors 1 (dashed) and 4 (solid). In the figures (a-b), the profiles are normalized to the same total deposited energy in order to isolate the pitch angle effect. This is noticeable only for sector 1, with the most shallow B-field inclination angle of 4$^\circ$, where the finite pitch angle leads to a slightly more relaxed energy deposition profile with 20\% (a) to 50\% (b) lower energy density values in the upper layers. 
 
The energy density profiles that are depicted in the lower row (c-d), are normalized according to the full-panel simulations which were discussed earlier. As aforementioned, the curvature and 3D effects result in significant differences between panels 1 and 4, as can be seen by comparing the solid and dashed curves in (c-d). The introduction of a finite pitch angle promotes further redistribution of the energy, increasing the energy deposited in sector 1 ($\sim$15\%) while decreasing that delivered to sector 4 ($\sim$20\%), which however still remains the most damaged in the case of 10\,MeV. In case of 50\,MeV, the introduction of 10$^\circ$ initial pitch combined with the highly relativistic nature of the REs leads to more pronounced redistribution, with sector 1 receiving approximately 40\% more energy and sector 4 about 10\% less compared to the zero pitch angle case.

\begin{figure*}
    \centering
    \subfloat{%
      \begin{overpic}[width =0.95\linewidth]{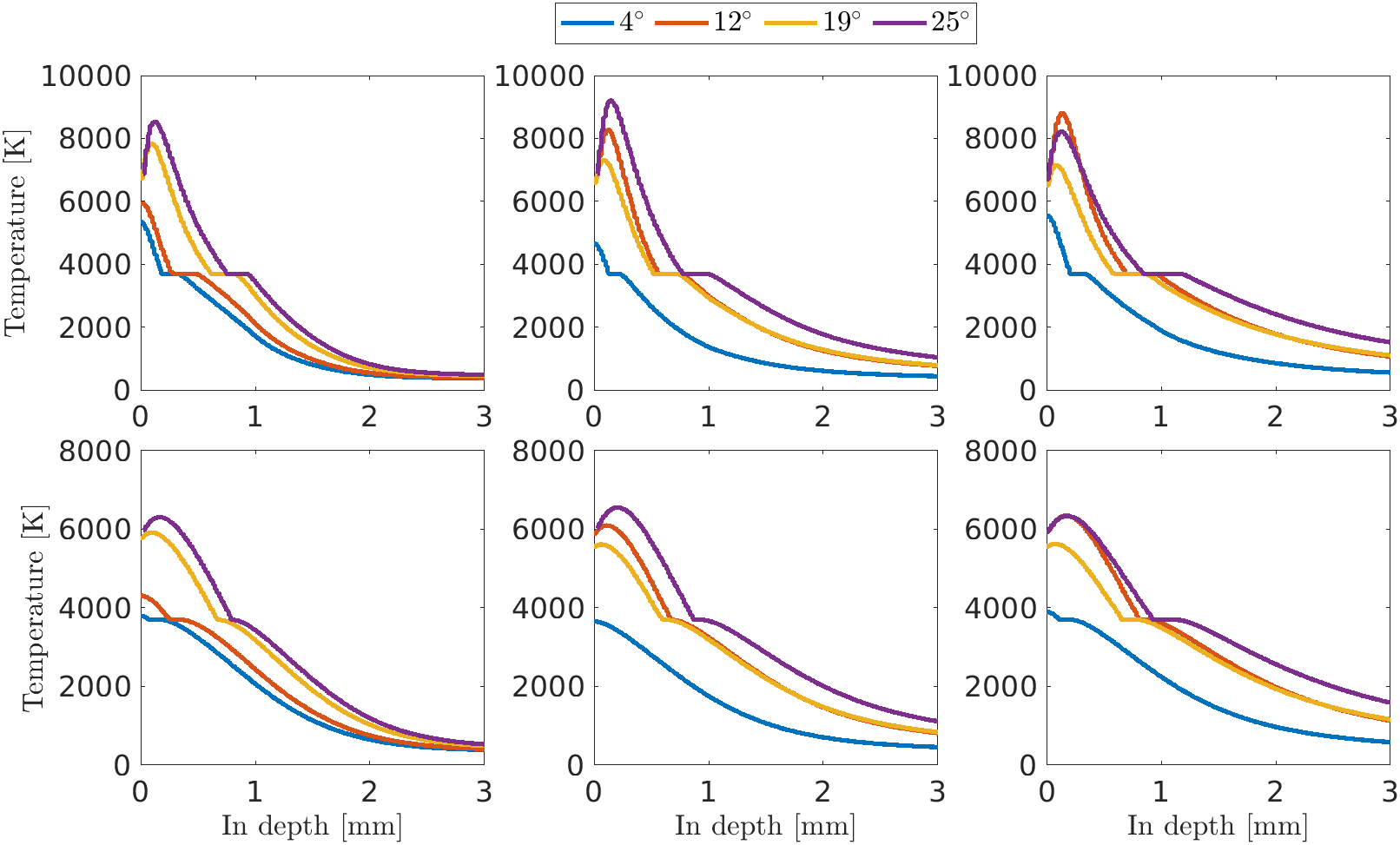}
          \put(29,52){\hbox{\kern3pt\textcolor{black}{\textbf{(a)} }}}
          \put(61,52){\hbox{\kern3pt\textcolor{black}{\textbf{(b)} }}}
          \put(94,52){\hbox{\kern3pt\textcolor{black}{\textbf{(c)} }}}
          \put(29,25){\hbox{\kern3pt\textcolor{black}{\textbf{(d)} }}}
          \put(61,25){\hbox{\kern3pt\textcolor{black}{\textbf{(e)} }}}
          \put(94,25){\hbox{\kern3pt\textcolor{black}{\textbf{(f)} }}}
      \end{overpic}          
    }
    \caption{MEMENTO results for the loading of 100\,kJ over 1\,ms (top row) and over 10\,ms (bottom row). From left to right: mono-energetic 10\,MeV\,\&\,10$^\circ$ pitch\, (a-d), \DREAM{} distribution\,I\, (b-e), \DREAM{} distribution\,II\, (c-f).}
    \label{fig:T_profiles}
\end{figure*}

\subsubsection{Thermal response of the panel.}

The thermal response is directly connected to the energy density maps. As discussed in Ref.\cite{Ratynskaia_2025b}, even monotonic in-depth energy density profiles can result in non-monotonic temperature distributions, with a maximum peaking underneath the surface. This is a consequence of the crucial role of the cooling flux due to vaporization. It is realized, for instance, in the case of 10 MeV\,\&\,100 kJ, where the mesh adopted is not fine enough to properly resolve the maximum energy density; however, a well pronounced peak is present in the temperature profile for sectors 3 and 4, as shown in Fig.\ref{fig:T_profiles} (a) and (d). On the other hand, for sufficiently high kinetic energies and steep impact angles, a non-monotonic temperature profile is a reflection of a non-monotonic energy density. This is for instance the case of 50 MeV REs, in sectors 3 and 4 [see Fig.\ref{fig:energy_densities}(d) for reference to the energy density profile]. Non-monotonic temperature profiles can lead to PFC explosions with debris release, as observed in multiple experiments, for details see Refs.\cite{Ratynskaia_2025b, Ratynskaia_2025, De_Angeli_2023, Ratynskaia_2025c}.

\begin{figure*}[!h]
    \centering
    \begin{subfigure}[t]{0.49\textwidth}
        \centering
        \includegraphics[width=\linewidth]{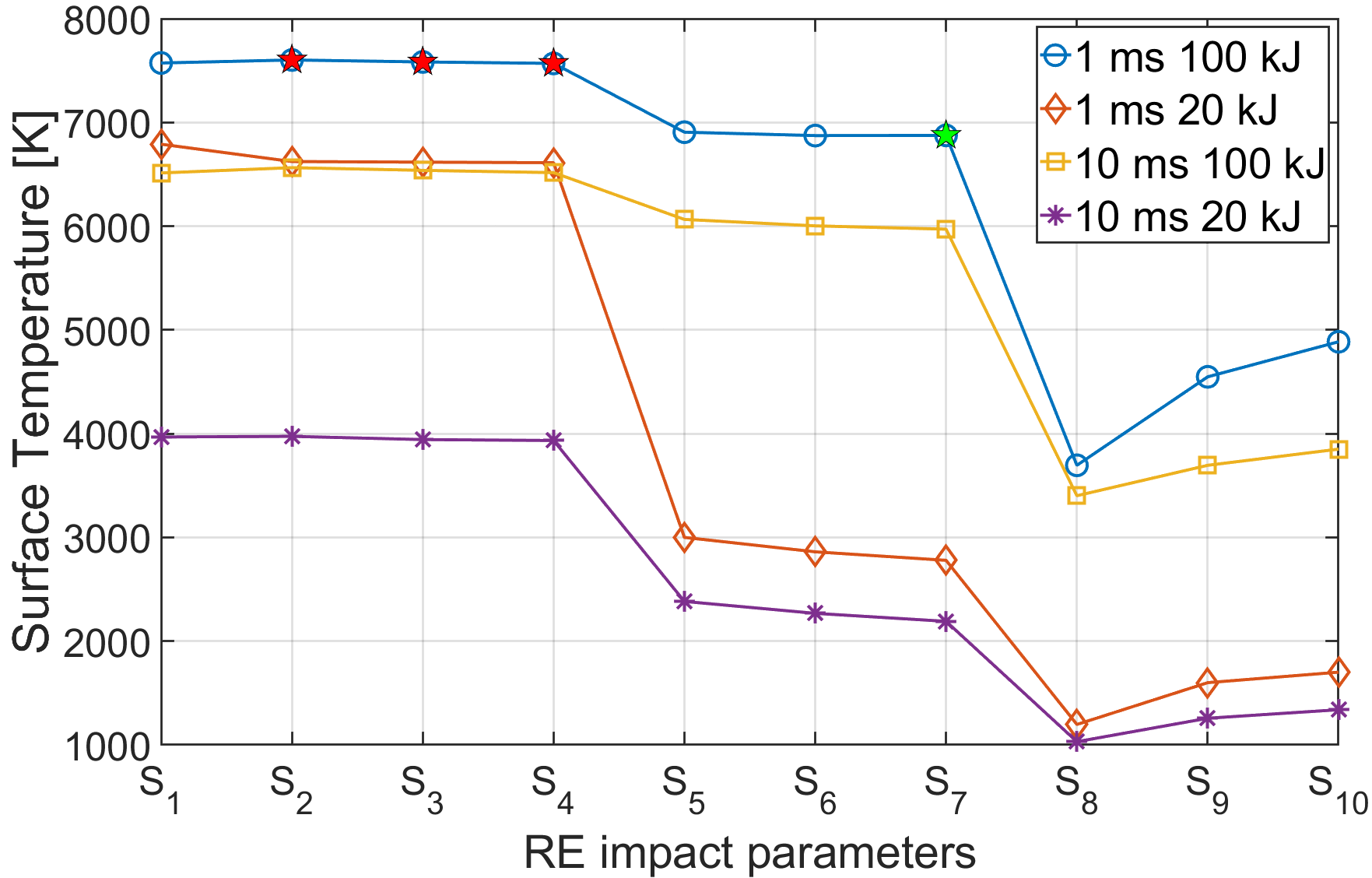}
        \caption{}
        \label{fig:T_surf}
    \end{subfigure}\hfill
    \begin{subfigure}[t]{0.49\textwidth}
        \centering
        \includegraphics[width=\linewidth]{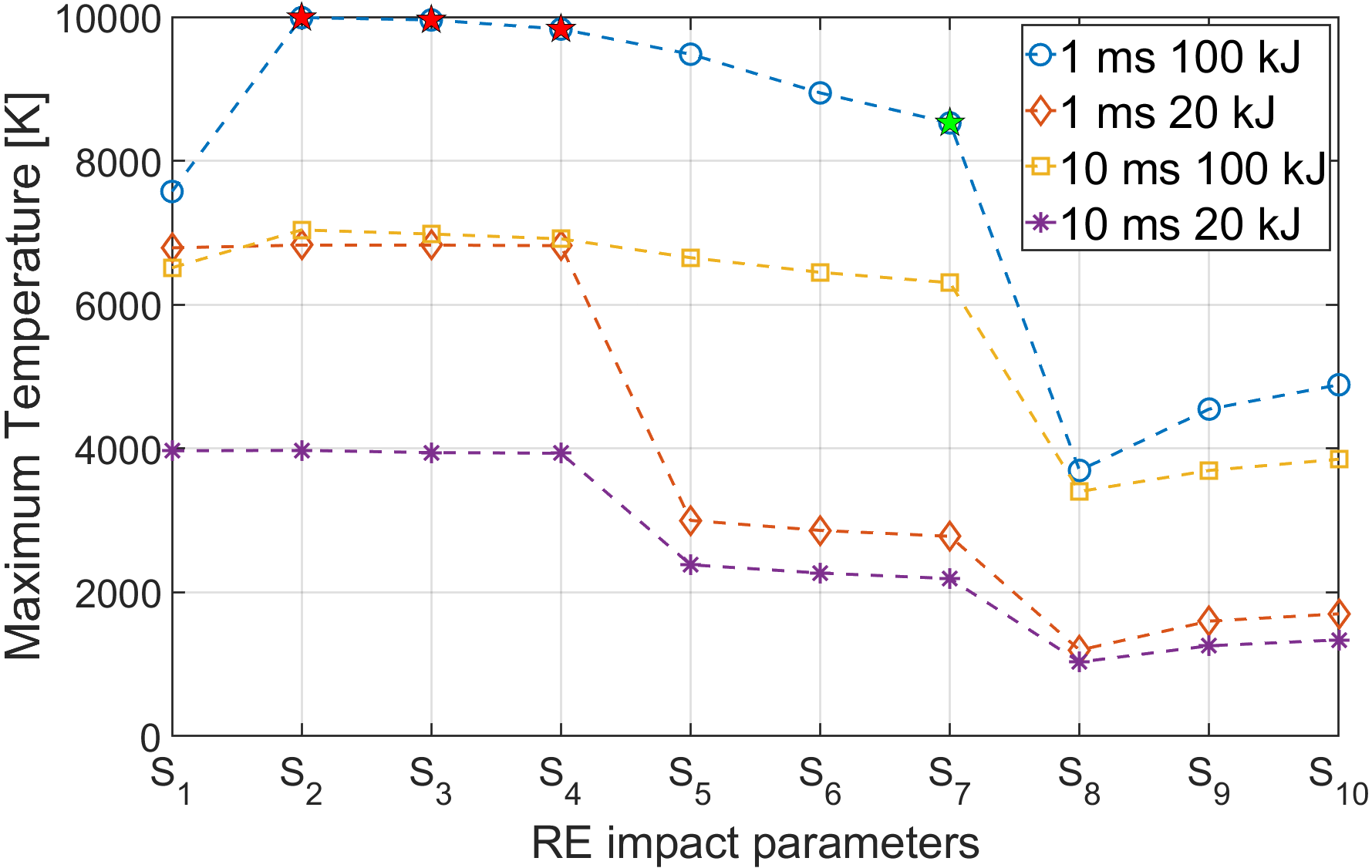}
        \caption{}
        \label{fig:T_max}
    \end{subfigure}
    \caption{Results of the Geant4-MEMENTO workflow for the worst-afflicted sector of each scenario, according to Table\,\ref{tab:scenarios}. The surface temperature $T_{\text{surf}}$ (a) and the maximum temperature reached $T_{\text{max}}$ (b) for the simplified input simulations. Note that simulations with 1\,MeV, marked with a red star, crashed after 0.2\,ms due to the temperature exceeding the validity range of the W thermophysical properties. The full temperature profiles for the scenario marked with a green star are shown in Fig.\ref{fig:T_profiles}.}
    \label{fig:T_combined}
\end{figure*}
\begin{figure}[!h]
    \centering
    \includegraphics[width=0.59\linewidth]{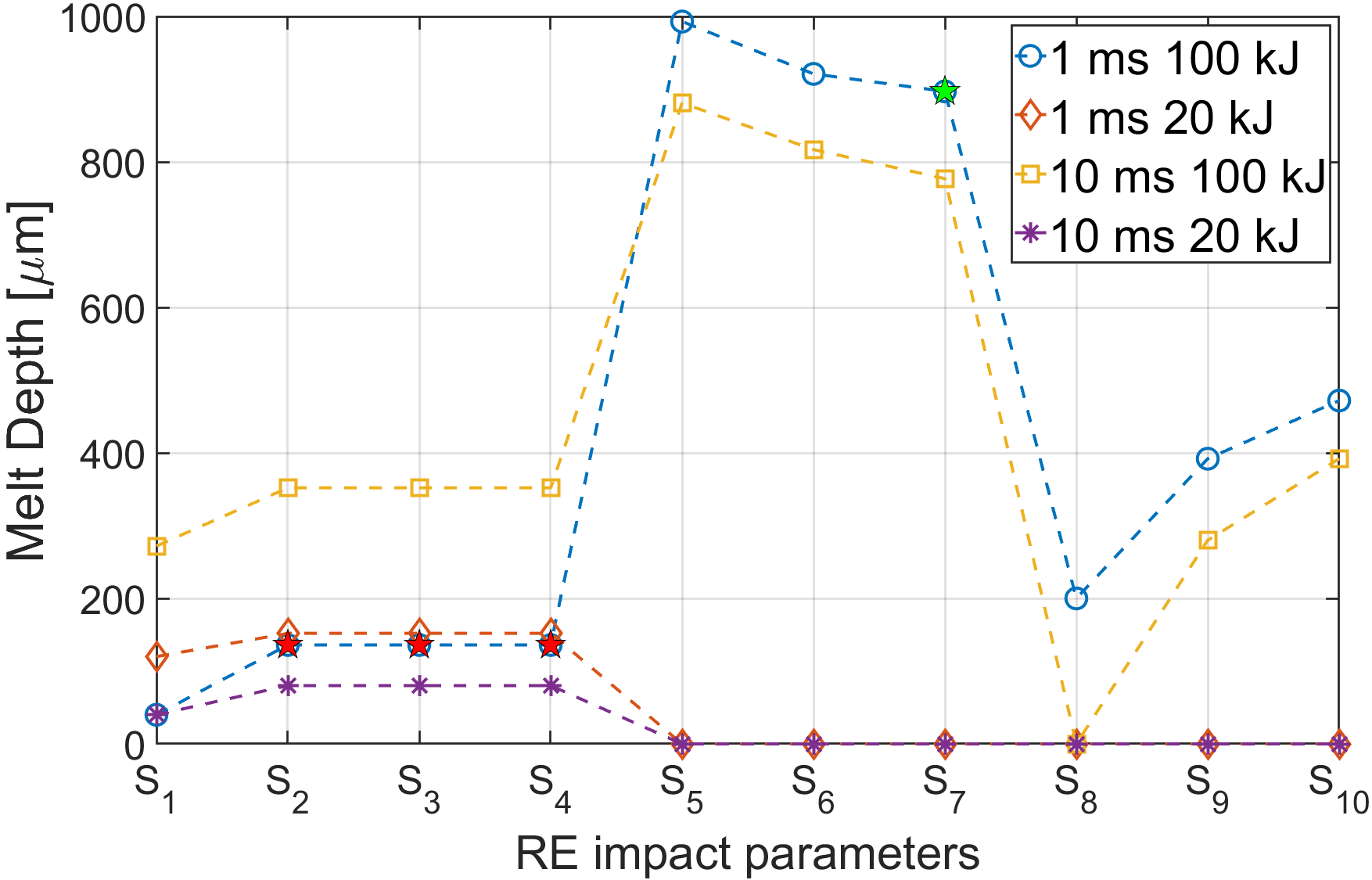}
    \caption{Results of the Geant4-MEMENTO workflow for the worst-afflicted sector of each scenario, according to Table\,\ref{tab:scenarios}. Plot of melt depth $h_{\text{melt}}$ for the simplified input simulations. Note that simulations with 1\,MeV, marked with a red star, crashed after 0.2\,ms due to the temperature exceeding the validity range of the W thermophysical properties. The full temperature profiles for the scenario marked with a green star are shown in Fig.\ref{fig:T_profiles}.}
    \label{fig:melt_depth}
\end{figure}

Figs.\,\ref{fig:T_combined} and \ref{fig:melt_depth} summarize the results for the most damaged sector of each scenario, i.e., where the highest energy densities and thus the most severe damage were observed. The RE impact energy-pitch combinations are labeled as scenarios $S_i$, as detailed in Table\,\ref{tab:scenarios}.

Comparing the surface temperature with the maximum temperature reached (Fig.\ref{fig:T_surf} and Fig.\ref{fig:T_max}), one can identify scenarios with non-monotonic temperature profiles. For 0.5\,MeV, the response is always monotonic, since this is essentially a case of surface loading. For 50\,MeV, on the other hand, the temperature profile is always non-monotonic, reflecting the energy deposition. In the case of 1\,MeV and 10\,MeV REs, whenever the surface temperature reached is sufficiently high to ensure non-negligible vaporization cooling, the profile becomes non-monotonic, as e.g. in the cases of 100\,kJ loaded over 1\,ms (10\,MeV and 1\,MeV) and of 20\,kJ loaded over 1\,ms (1\,MeV). 

Naturally, lower electron energies result in higher energy densities, and shorter loading times result in higher power densities and, thus, higher temperatures near the surface. These trends can be observed in the plots of Fig.\ref{fig:T_surf} and Fig.\ref{fig:T_max}. 

In all simulations with an initial RE energy $\leq$10\,MeV, the melting is deepest in sector 4, which not only receives the highest energy (see Section \ref{sec:E_distr}) but also has the largest impact angles (due to highest $\alpha$) leading to less steep energy deposition profiles. As seen from Fig.\ref{fig:melt_depth}, at the lower limit of 20\,kJ deposited, melting occurs only for the 1\,MeV and 0.5\,MeV cases. Higher electron energies lead to more relaxed energy density deposition maps, while the temperature profiles are relaxed by diffusion during longer loading times. Indeed, for 100\,kJ loading, the 10\,MeV cases (S5-S7) show the highest, $\sim$\,1\,mm, melt depths. In the cases of 50\,MeV (S8-S10), the highest energy densities are in sectors 1 and 2, where, in spite of shallow $\alpha$ and hence impact angles, melt depths up to $\sim 0.5$\,mm are found.


Strong vaporization makes the 0.5\,MeV scenario stand out from other runs, with $\sim$\SI{350}{\micro\meter} eroded in 1\,ms, for the scenario of 100\,kJ total energy deposited, and with more than $\sim$55\% of the energy dissipated by vaporization (for the sake of comparison, about $\sim$15\% of the energy is dissipated by vaporization in the same scenario for the 1\,MeV REs). The loading is intense enough to promote a high vaporization regime, where the erosion of the top surface progresses rapidly, effectively competing with the melt generation. As a consequence, although the highest temperatures are recorded in sector 4, the deepest melting is encountered in different sectors in each of the four corresponding scenarios. When loading is distributed over 10\,ms, only low RE energies (0.5 and 1\,MeV) result in appreciable, $\sim$\SI{200}{\micro\meter},  vaporization. On the other hand, the 20\,kJ loading leads to none or negligible, $<$\SI{50}{\micro\meter}, erosion.

The effect of a non-zero pitch angle on the thermal response is negligible for 1\,MeV REs. To appreciate differences in the in-depth energy deposition profile for the low energy scenario of 1\,MeV, a resolution down to tens of nanometers or even nanometers is required at grazing impact angles\,\cite{Ratynskaia_2025b, Ratynskaia_2025c}. For higher initial energies, the effect is masked by the way the deposited energy is re-distributed across the curved tile, as described in Section \ref{sec:E_dens_distr}. As a reflection of the energy deposition profiles in Fig.\ref{fig:energy_densities}(c), where the solid curves (sector 4) correspond to scenarios S5 \& S7, we find that a non-zero pitch yields lower maximum temperature (Fig.\ref{fig:T_max}) and melt depth (Fig.\ref{fig:melt_depth}). In contrast, for the Fig.\ref{fig:energy_densities}(d) profiles, where the blue dashed curve (sector 1) corresponds to S10, a non-zero pitch yields higher maximum temperature and melt depth.

\subsection{Simulations for DREAM distribution functions}

\subsubsection{Energy distribution on the panel.}

The energy partitioning between the four sectors lies far from the pure geometrical $\sin\alpha$ scaling, primarily owing to the introduction of a non-zero pitch angle, extending up to 70$^\circ$ (${\sim}1.2~\mathrm{rad}$), but also because of the high energy tails of the \DREAM{} distributions [up to 60\,MeV in scenario (I) and up to 100\,MeV in (II)], as explained in Section \ref{sec:E_distr}. As a consequence, sector 4 receives almost three times the energy of sector 1, while sector 2 and sector 3 receive approximately the same energy, i.e. twice what received by sector 1.

\subsubsection{Energy density distribution on the panel-}

As already described in Section \ref{sec:E_dens_distr}, the highest energy density is expected in sector 1, characterized by the most shallow incidence of the magnetic field. However, this sector also receives the lowest total energy, as discussed above, and the highest energy density is hence encountered in sector 2, for both scenarios.

\begin{table}
    \centering
    \resizebox{\columnwidth}{!}{%
    \begin{tabular}{ c | l l l l l l l l l l  }
         \hline
         &\cellcolor{lightgray}$S_1$
         &\cellcolor{lightgray}$S_2$
         &\cellcolor{lightgray}$S_3$
         &\cellcolor{lightgray}$S_4$
         &\cellcolor{lightgray}$S_5$
         &\cellcolor{lightgray}$S_6$
         &\cellcolor{lightgray}$S_7$
         &\cellcolor{lightgray}$S_8$
         &\cellcolor{lightgray}$S_9$
         &\cellcolor{lightgray}$S_{10}$\\
         \hline
      $E_k$ (MeV)   & 0.5& 1& 1& 1& 10&10&10&50&50&50\\ 
      \hline
      Pitch angle (deg) &0& 0&5&10&0&5&10&0&5&10\\
      \hline
      Sector &4&4&4&4&4&4&4&2&1&1\\
      \hline
    \end{tabular}%
    }
    \caption{Incident RE kinetic energy and pitch angle for the scenarios $S_i$ referred to in Figs.\ref{fig:T_surf}-\ref{fig:melt_depth}. The third row designates the most damaged sector in each scenario.}
    \label{tab:scenarios}
\end{table}

\subsubsection{Thermal response of the panel.}

The temperature profiles for each of the four sectors are reported in Fig.\,\ref{fig:T_profiles} enabling comparison between the two loading times of 1\,ms and 10\,ms, see the top and bottom rows respectively. For comparison with the simplified input simulations, the case of 10\,MeV and 10$^\circ$ pitch was chosen since it closely represents the most probable part of both \DREAM{} distributions; see the white-yellow regions in Fig.\ref{fig:distr_DREAM} (a-b). For sectors 3 and 4, no substantial differences are observed between the \DREAM{} distributions and the mono-energetic case, as shown by the purple curves. Similar surface and maximum temperatures are reached. Simulations adopting \DREAM{} distributions lead to a melt depth slightly deeper in sector 4 ($\sim10\%$ for scenario I and $\sim20\%$ for scenario II), but slightly shallower in sector 3 ($\sim15\%$ for scenario I and $\sim20\%$ for scenario II).

On the other hand, for sector 2, the monoenergetic case results are characterized by a nearly a factor of two discrepancy in the melt depth as well as a 30$\%$ higher maximum temperature compared to the \DREAM{} distribution results, see the red curves in Fig.\ref{fig:T_profiles}. This is mostly the effect of the high energy tail setting apart both \DREAM{} distributions from the monoenergetic 10\,MeV case, so that the energy partition over the curved panel geometry deviates significantly from the pure geometric projection. 

\begin{table*}[!t]
    \centering
    \setlength{\tabcolsep}{3pt}
    \begin{tabular}{l c c c c c}     
        \textbf{Case Reference} & \textbf{\% energy loss} &  \textbf{EBS } & \textbf{PBS } & \textbf{Photons} & \textbf{Neutrons} \\ 
        \hline \hline
        $S_1$ (0.5\,MeV\,\&\,0$^\circ$\,pitch)      & 5.70\%  & 5.22\%  & -         & 0.48\% & -  \\
        $S_2$ (1\,MeV\,\&\,0$^\circ$\,pitch)         & 12.85\% & 11.16\% & -         & 1.69\% & -   \\
        $S_3$ (1\,MeV\,\&\,5$^\circ$\,pitch)  & 12.76\% & 11.08\% & -         & 1.68\% & -   \\
        $S_4$ (1\,MeV\,\&\,10$^\circ$\,pitch)  & 12.31\% & 10.65\% & -         & 1.66\% & -   \\
        $S_5$ (10\,MeV\,\&\,0$^\circ$\,pitch)       & 18.82\% & 6.10\%  & 0.0084\%  & 12.71\% &2.39E-5\%   \\
        $S_6$ (10\,MeV\,\&\,5$^\circ$\,pitch)  & 17.88\% & 4.57\%  & 0.0091\%  & 13.30\% & 1.76E-5\%  \\
        $S_7$ (10\,MeV\,\&\,10$^\circ$\,pitch)  & 17.26\% & 4.46\%  & 0.0082\%  & 12.79\% &1.13E-5\%   \\
        $S_8$ (50\,MeV\,\&\,0$^\circ$\,pitch)       & 27.27\% & 8.11\%  & 0.65\%    & 18.30\% &  0.21\% \\
        $S_9$ (50\,MeV\,\&\,5$^\circ$\,pitch)  & 22.41\% & 3.21\%  & 0.13\%    & 18.84\% & 0.23\%  \\
        $S_{10}$ (50\,MeV\,\&\,10$^\circ$\,pitch)    & 21.06\% & 3.01\%  & 0.14\%    & 17.70\% & 0.21\%  \\ 
        \DREAM{} distribution I  &       19.21\%      &  3.76\%      &       0.06\%     &     15.38\% &  0.01\%\\
        \DREAM{} distribution II  &        20.09\%     &     3.99\%   &         0.09\%   &      15.99\% & 0.02\% \\
        \hline \hline
    \end{tabular}
    \caption{Results of the full panel Geant4 simulations for the energy losses due to particle backscattering or transmission, expressed as a percentage of the total energy loaded, Here, EBS stands for Electron Backscattering and PBS for Positron Backscattering. Note that neither neutrons nor positrons are produced in the low energy scenarios, since the electron energy lies below the threshold of the cross-sections for photoneutron production and gamma conversion, respectively.}
    \label{tab:energy_loss}
\end{table*}

\subsection{Evaluation of particle loss channels}

As commented in Section \ref{sec:experiment}, the full panel simulations facilitate estimates of energy losses due to primary or secondary particles being backscattered or being transmitted through the sample volume. To be more specific, the simulations account for \textbf{(i)} electron backscattering (EBS) of primary or secondary electrons, \textbf{(ii)} positron backscattering (PBS) of secondary positrons, \textbf{(iii)} photon transmission losses, \textbf{(iv)} neutron transmission losses.

Table \ref{tab:energy_loss} reports the total energy loss as a percentage of the energy loaded for every simulated scenario and its subdivision into the four dissipation channels. It is clear that the total energy loss increases with the initial energy of the RE population: from  $\sim$5\% at 0.5\,MeV to more than 20\% at 50\,MeV, while the \DREAM{} scenario losses are in between the 10\,MeV and the 50\,MeV mono-energetic cases, as expected from the most probable part of the \DREAM{} distributions. The reason is related to the bremsstrahlung contribution, the higher the primary electron energy the larger the energy portion converted into photons, which more easily escape the panel volume. To be more specific, starting from incident energies roughly exceeding 10.5\,MeV in W, Bremsstrahlung is the primary electron energy dissipation mechanism overcoming ionization-excitation losses\,\cite{Berger_1970,Berger_1992}. 

The trend for energy losses due to backscattering is opposite, since the backscattering yield strongly decreases with the energy of the primary electron in the relativistic regime. Note that electron backscattering is enhanced at grazing angles, but a fraction of the  backscattered electrons can be re-deposited due to gyromotion. 

Essentially, neither neutrons nor positrons are generated for low energy simulations (0.5 and 1\,MeV) while the energy losses are negligible ($<1$\%) for higher RE energies. Neutrons originate primarily from photo-nuclear reactions, with a kinematic threshold of about 8\,MeV for W\,\cite{HandbookIAEA_2000,Kawano2020}, representing the rarest particles produced in terms of cross-sections. 

The relative uncertainty in neutron production is approximately 2\%. On the other hand, the uncertainty in photon production is negligible across the considered energy range owing to the high incident electron statistics. The photons are predominantly generated via bremsstrahlung (with additional contributions from atomic de-excitation, Compton scattering, and electron–positron annihilation) which ensures statistically robust photon statistics. The positrons are mainly produced through pair production, which has a threshold of 1.022\,MeV and becomes the dominant photon interaction process in W above 10\,MeV. Consequently, at both 10\,MeV and 50\,MeV, the relative uncertainty in positron production is negligible. However, the majority of positrons annihilate with valence or core electrons during transport, resulting in a much lower escape fraction from the panel.

\section{Conclusions and Outlook}

Modeling of the thermal response of a realistic PFC tile geometry from SPARC outboard off-midplane limiters to runaway electron impacts has been carried out. Parametric scans of impacting RE characteristics as well as energy-pitch distribution functions simulated by \DREAM{} have been employed for the volumetric energy source calculations assuming deposition time scales of 1 and 10\,ms. Our Monte Carlo simulations of the energy deposition reveal the crucial role of the 3D effects. In particular, modeling of a realistic curved geometry to account for the local magnetic field inclination angle $\alpha$ and modeling of the re-deposition of backscattered electrons are required for accurate estimates of energy partition over the tile. The smearing out effect, more pronounced at higher RE energies and larger Larmor radii, makes the energy partition over the curved panel deviate from the simple `optical' loading proportional to $\sin \alpha$. This is also the reason why characterization of impacting REs with a single pitch-energy combination - even if chosen from the most probable part of the RE distribution - does not yield similar predictions for the thermal response. The high energy tail will deposit energy highly nonlocally and change the energy partition over the panel and hence the location of the most damaged part.

For incident RE energies of $10$\,MeV and above, the loading of 20\,kJ does not result in any damage to the limiter, since the temperatures do not reach the melting point. However, if such loading is provided by low energy ($\sim$ 1\,MeV) REs, melting up to $\sim$\SI{100}{\micro\meter} can be induced. With 100\,kJ loaded, RE energies of 10\,MeV and above lead to appreciable, from a few hundred micrometers up to a millimeter, melt depths. Moreover, these scenarios are characterized by non-monotonic temperature profiles which could initiate explosive responses accompanied by the violent release of debris\,\cite{Ratynskaia_2025b, Ratynskaia_2025, De_Angeli_2023,Ratynskaia_2025c}. For 100\,kJ over 10\,ms loading, low ($\sim$ 1\,MeV) energy RE impacts result in strong, up to $\sim$\SI{500}{\micro\meter}, vaporization losses. The same loading but over 1\,ms cannot be fully simulated due to reaching extreme surface temperatures where the adapted W thermophysical properties are not valid.

This study presents an important starting point for the prediction, understanding, and possibly even mitigation of the thermal damage from RE impacts in SPARC. In particular, the modeling results highlight some key subtleties of the energy deposition, melting and vaporization from plausible RE distributions with realistic PFC and B-field geometries. For a complete picture of material damage and surface topology after RE-induced explosive events, modeling of the full thermomechanical response is required which comprises a challenging problem\,\cite{Ratynskaia_2025c, Ratynskaia_2025b}. Any follow-up studies will need to account for such likelihood and risk as part of the disruption ``budget''. Furthermore, it must be evaluated how energy transmission (e.g. over 10\% via bremsstrahlung) will impact the surrounding structural materials, with lower melting temperatures, or even the superconducting magnets.

In parallel, work is ongoing to refine the RE loading input into such thermal response workflows: REs as a fluid are being modeled self-consistently with 3D nonlinear MHD with the M3D-C1 code, including updated VDE simulations\,\cite{Datta2025,Datta2026}, while complementary \DREAM{} simulations are further improving kinetic descriptions of REs in SPARC mitigated by massive material injection\,\cite{Ekmark_2025}. Individual particle following is being advanced within M3D-C1 itself and being extended to full SPARC CAD geometries within the HEAT code\,\cite{Looby_2022_FST,Looby_2022_NF,Feyrer_2026}. A scan of the RE impacts, as performed in this paper, serves as a strong basis for future surrogate modeling, as well as an opportunity for experimental validation with a data set of MeV-range electron impacts on pure W samples carried out with a Van de Graaff generator. Lastly and perhaps most importantly, this work indicates that many of the lessons learned on SPARC will inform RE impacts on ARC, its planned successor and future fusion pilot plant\,\cite{Sweeney_2026}.

\section*{Acknowledgments}

The Geant4 and MEMENTO simulations were enabled by resources provided by the National Academic Infrastructure for Supercomputing in Sweden (NAISS) at the NSC (Link\"oping University) partially funded by the Swedish Research Council through grant agreement No\,2022-06725. The kinetic \DREAM{} simulations were enabled by the support of the Swedish Research Council (Dnr.~2022-02862 and 2021-03943), and by the Knut and Alice Wallenberg foundation. 

\noindent The authors are grateful to contributions from C.~Rea, C.~Clauser, and R.~Sweeney. This work was funded in part by Commonwealth Fusion Systems.

\section*{References}
\noindent  
\bibliography{biblio_new}
\end{document}